MAGNETIC EFFECTS AND OVERSIZED M DWARFS IN THE YOUNG OPEN CLUSTER NGC 2516

James MacDonald and Dermott J. Mullan

Dept. of Physics and Astronomy, University of Delaware, Newark, DE 19716

ABSTRACT

By combining rotation periods with spectroscopic determinations of projected rotation velocity, Jackson, Jeffries & Maxted (2009) have found that the mean radii for low-mass M-dwarfs in the young, open cluster NGC 2516 are larger than model predictions at a given absolute *I* magnitude or I - *K* color and also larger than measured radii of magnetically inactive M-dwarfs. The relative radius difference is correlated with magnitude, increasing from a few per cent at $M_I = 7$ to greater than 50 per cent for the lowest luminosity stars in their sample at $M_I \sim 9.5$. Jackson et al (2009) have suggested that a two-temperature star spot model is capable of explaining the observations, but their model requires spot coverage fractions of at least 50 per cent in rapidly rotating M-dwarfs. Here we examine these results in terms of stellar models that include the inhibiting effects of magnetic fields on convective energy transport, with and without the effects of star spots. We find that a pure spot model is inconsistent with the color – magnitude diagram. The observations of radii versus color and radii versus absolute magnitude in NGC 2516 are consistent with models which include only magnetic inhibition or a combination of magnetic inhibition and spots. At a given mass we find a large dispersion in the strength of the vertical component of the magnetic field in the stellar photosphere but the general trend is that the vertical field increases with decreasing mass from a few hundred Gauss at 0.65 $M_\odot$ to 600 - 900 Gauss, depending on spot coverage, in the lowest mass stars in the sample at 0.25 $M_\odot$.

1. INTRODUCTION

A long term problem with modeling lower main sequence stars is that a growing number of objects have observationally determined radii that are significantly larger than model predictions ((López-Morales 2007; Ribas et al. 2008). In many cases, this oversizing is associated with magnetic activity signatures (Torres & Ribas 2002; Morales et al. 2009 and references therein). A classic example is CM Draconis (CM Dra), an eclipsing binary containing two dM4.5 stars with masses of 0.23102 ± 0.00089 and 0.21409 ± 0.00083 $M_\odot$, and empirical radii of 0.2534 ± 0.0019 and 0.2398 ± 0.0018 $R_\odot$, respectively (Morales et al. 2009; Torres et al. 2010). Comparing these radii with stellar models having a range of ages and heavy element abundances, it emerges that both components have radii which are larger than main-sequence models predict by at least 0.01 $R_\odot$ (Morales et al. 2009). In view of the small statistical errors that are quoted for the empirical radii, this oversizing of the radii of both components is at least a $5\sigma$ effect. CM Dra is a close binary with a short orbital period of 1.27 d. Based on the age determination of a proper motion white dwarf companion, CM Dra is sufficiently old at 4 Gyr that its components are likely to be in synchronous rotation with the orbit. This short rotation period is often used as an argument for strong dynamo action. Indeed CM Dra shows evidence for magnetic fields in that it has spots and flares (this evidence has been recently reviewed by MacDonald & Mullan 2012). However, oversized stars do occur in binary systems with significantly longer periods. A recently discovered example is LSPM J1112+7626, which is a M-dwarf eclipsing binary with a 41 d orbital period (Irwin et al 2011). The component masses and radii are $M_1 = 0.395 \pm 0.002$ $M_\odot$, $R_1 = 0.3860 \pm^{0.0055}_{0.0052}$ $R_\odot$, and $M_2 = 0.275 \pm 0.001$ $M_\odot$, $R_2 = 0.2978 \pm^{0.0049}_{0.0046}$ $R_\odot$. A 65 day out-of-eclipse modulation is attributed to rotational modulation of photospheric spots on one of the binary components. Compared to Baraffe et al. (1998) models for 10 Gyr age, [M/H] = 0, the observed radii are 3% and 5% larger. It is particularly of note that the two stars straddle the boundary between partially radiative and fully convective and that the relative oversizing is larger in the fully convective secondary. Detailed modeling of LSPM J1112 is hampered by lack of good age and metallicity estimates. Irwin et al (2011) note that i) the presence of strong TiO bands indicates that LSPM J1112 is not extremely metal poor; they estimate [Fe/H] > −1 and the heavy element abundance is probably closer to solar (or greater), and ii) although the radii are consistent with an age of 120 Myr for solar metallicity, a low age seems ruled out by the old disk kinematics, and the lack of strong H$\alpha$ emission and evidence for rapid rotation.

Some of the issues of age and metallicity could be avoided by studying M dwarf stars in clusters. Jackson, Jeffries & Maxted (2009, hereafter J09) have combined spectroscopic determinations of projected rotation velocity with photometric determinations of the rotational period (Irwin et al. 2007) for low mass stars in the open cluster NGC2516. Spectroscopic and photometric determinations indicate a near solar composition (Sung et al. 2002; Terndrup et al. 2002) and, from the main sequence turnoff, the age is determined to be 125 ± 25 Myr (Lyra et al. 2006). J09 find that the radii of single, rapidly rotating M-dwarfs in NGC 2516 are significantly larger than predicted by evolutionary models. The discrepancy is larger for stars with lower luminosity, and largest for stars of spectral type which are predicted to be fully convective, reaching a radius anomaly of greater than 50 per cent. J09 show that a simple two-temperature star spot model could explain the overall radius increase, but only if spots cover ≥ 50 per cent of the surface for the lowest luminosity stars.

In this paper, we consider the results of J09 in the context of a model for magneto-convection, with and without inclusion of the effects of star spots. For comparison with the J09 results, we use their adopted age for NGC 2516 of 160 Myr, rather than the slightly lower age found by Lyra et al (2006). Since low mass pre-main sequence stars contract at almost constant effective temperature, $T_{eff}$, age uncertainties have a negligible effect on the radius – absolute magnitude and radius – color diagrams that we use to compare our results with those of J09.

## 2. MODELING TECHNIQUE

### 2.1. Input physics

The current version of our stellar evolution code has a number of options for equation of state (eos) and treatment of outer boundary condition. As discussed in MacDonald & Mullan (2012), the choice of eos has only a few percent effect on the radii of fully convective main sequence stars. Since NGC 2516 is relatively young, stars of mass less than 0.5 $M_\odot$ have not fully settled on the main sequence. The 160 Myr solar composition radius – mass isochrone is shown in figure 1a for a few different choices of eos. The OPAL eos is described by Rogers & Nayfonov (2002) and the SCVH eos is the Saumon, Chabrier & van Horn (1995) eos for hydrogen – helium mixtures. The SCVH + Z eos is our modification of the SCVH eos to include a correction for finite heavy element abundance. For three of the isochrones shown in figure 1a, we have used NextGen atmospheres (Hauschildt, Allard, & Baron 1999) to determine the outer boundary conditions. Specifically, we use the temperature and pressure at optical depth 100, for given $T_{eff}$, surface gravity and composition. Since these atmosphere models use mixing length ratio, $\alpha = 1$, we have used $\alpha = 1$ in the interior for all models. We find that the OPAL eos gives larger radii than the SCVH + Z eos by at most 1.5%. This difference is significantly less than the up to 50% oversizing found by J09 and our conclusions will not be seriously affected by our choice of eos. Since the OPAL eos gives the larger radii, we have elected to use the OPAL eos. Since the NextGen atmospheres use standard mixing length theory for convection, it is not consistent to use them for models that use our magneto-convection prescription. We therefore determine the outer boundary condition by using the Eddington approximation to determine the temperature and pressure at optical depth 0.1. The radii of non-magnetic models calculated using the Eddington boundary condition are also compared with those calculated with the atmosphere boundary condition in figure 1a. We see that the atmosphere boundary condition models are larger by 5.7% for models of mass 0.15 $M_\odot$. The radius difference decreases with increasing mass to 1.5% at 0.70 $M_\odot$. Again these differences are relatively small compared to the observed oversizing. Figure 1b shows the $T_{eff}$ – mass isochrones at the same age. We see that the models using the Eddington boundary condition have higher $T_{eff}$ than the models using the atmosphere boundary condition, with the maximum temperature difference of 160 K occurring for the 0.15 $M_\odot$ model. For models with mass between 0.4 and 0.6 M$_\odot$, the temperature difference is about 100 K. These temperature differences have a small but not insignificant effect on the colors of the models, which we will discuss later. From now on we make use only of models calculated using the OPAL eos and the Eddington boundary condition.

*2.2. Magneto-convection*

Our model for magneto-convection has been developed in a sequence of papers beginning with Mullan & MacDonald (2001). It is based on a criterion derived by Gough & Tayler (1966) who found that convective stability in the presence of a vertical magnetic field $B_v$ frozen in an ideal gas of pressure $P_{gas}$ and infinite electrical conductivity is ensured as long as $\nabla_{rad}$ does not exceed $\nabla_{ad} + \delta$, where the magnetic inhibition parameter $\delta$ is defined (in Gaussian cgs units) by

$$\delta = \frac{B_v^2}{B_v^2 + 4\pi\gamma P_{gas}}, \qquad (1)$$

Here $\gamma$ is the first adiabatic exponent.

The Gough – Tayler criterion may not directly apply to convection in a cool magnetic dwarf for two reasons: 1) The gas is far from ideal and 2) the electrical resistivity can be much higher than for fully ionized plasma. The finite magnetic diffusivity allows fluid to move across magnetic field lines, and this can weaken the ability of the magnetic field to hinder thermal convection. To take into account non-ideal thermodynamic behavior and the effects of finite electrical conductivity, Mullan & MacDonald (2010) modified the Gough & Tayler criterion for convection to

$$\nabla_{rad} - \nabla_{ad} > \Delta \equiv \frac{\delta}{\theta_e} \min\left(1, \frac{2\pi^2 \gamma \kappa}{\eta \alpha^2}\right). \qquad (2)$$

Here the non-ideal behavior is accounted for by the inclusion of a dependence on the thermal expansion coefficient $\theta_e = -\partial \ln \rho / \partial \ln T\big|_P$. The factor $\min\left(1, 2\pi^2 \gamma \kappa / \eta \alpha^2\right)$, in which $\kappa$ is the thermal conductivity and $\eta$ is the magnetic diffusivity, gives a transition to a generalization (Cowling 1957) of the criterion derived by Chandrasekhar (1961, and references therein) for the onset of thermal convection in the presence of a vertical magnetic field in a thermally conducting and magnetically diffusive incompressible fluid. Our method of calculation of the electrical conductivity is given in MacDonald & Mullan (2009). We shall refer to eq. (0) as the GTC criterion when the finite conductivity correction factor is included and as the GT criterion when the finite conductivity correction is omitted.

Although the GTC criterion gives the correct qualitative behavior in the limits of $\kappa/\eta$ being large or small, we do not claim that it is precise. Uncertainties lie in the choice of the numerical factor $2\pi^2/\alpha^2$, and an additional multiplicative factor, $f_{ec}$, could be included to allow for these uncertainties in the effects of electrical conductivity. However, it is important to keep in mind the following two observational properties of the stars which are being discussed here: (i) spots are present (J09) and (ii) there is significant chromospheric activity (Jackson & Jeffries 2010) in all of the stars in the J09 sample. Both of these properties indicate clearly that there must be significant coupling between matter and magnetic field in the J09 stars. The existence of such coupling suggests that finite conductivity effects may be negligible. Hence in this work, we have used the GT criterion, i.e. eq. (0), without the finite conductivity correction factor, as the criterion for onset of convective stability.

To determine the convective energy flux, we replace $\nabla_{ad}$ by $\nabla_{ad}+\Delta$ everywhere it appears in the mixing length theory. Our specific choice of mixing length theory is that of Mihalas (1978), which is the same as that of Böhm-Vitense (1958) but modified to include a correction to radiative losses from convective elements when they are optically thin.

The magnetic inhibition parameter $\delta$ is a local variable: in general, its numerical value may vary as a function of radial position in a star, and the question arises as to the appropriate choice for the radial profile of $\delta(r)$. Here we make use of dynamo concepts, as reported in our modeling of CM Dra (MacDonald & Mullan 2012), to set the profile. We take $\delta$ to be constant from the surface to the radial location at which the local magnetic field strength reaches a prescribed value (the 'ceiling'). At deeper radial locations, the field is held fixed at its ceiling value which we take to be 1 MG.

*2.3. Star spots*

Our treatment of star spots is the same as in MacDonald & Mullan (2012). The influence of star spots on internal stellar structure has been reviewed by Spruit (1992). The blocking effects of spots is modeled by modifying the surface boundary condition to

$$L = 4\pi R^2 \left(1 - f_s\right) \sigma T_u^4, \tag{3}$$

where $f_s$ is the effective fraction of the surface covered by spots, which are assumed to be completely dark, and $T_u$ is the surface temperature of the immaculate (unspotted) surface, which is set equal to the model temperature at optical depth 2/3. For fully convective stars, spots result in a reduction in luminosity given by $\Delta L/L \approx -f_s$, with much smaller relative changes in $R$ and $T_u$. Note that because $T_u$ does not change significantly in the presence of a spot, any significant spot coverage will reduce $T_{eff}$ for fully convective stars according to the expression $T_{eff}^4 = (1 - f_s)T_u^4$. Hence to avoid confusion it is important to carefully distinguish between $T_u$ and $T_{eff}$ for heavily spotted stars. This point is particularly relevant to determining bolometric corrections and colors. For completely dark spots, the observed radiation comes totally from the unspotted surface and $T_u$ is the appropriate temperature to use in determining bolometric corrections and colors. If the spots are not completely dark, then the appropriate temperature to use will lie between $T_u$ and $T_{eff}$. Following J09, we have taken the spots to be 30% cooler than the rest of the surface i.e. $T_s = 0.7T_u$. The magnitude and color of the models are then determined by appropriately combining the magnitudes and colors of the two surfaces of areas $4\pi R^2 (1 - f_s)$ and $4\pi R^2 f_s$ with respective temperatures $T_u$ and $T_s$.

*2.4. Bolometric corrections and colors*

To compare our results with the observations, we need a way to convert from our theoretical $T_{eff}$ and log $g$ to the absolute Cousins *I* and CIT *K* magnitudes used by J09. We have considered i) the Johnson-Cousins-Glass system bolometric corrections computed by Bessell, Castelli & Plez (1998) from synthetic spectra, ii) for $I_C$, the empirically constrained bolometric corrections of VandenBerg & Clem (2003) calculated with subroutines kindly provided by Don VandenBerg, and iii) The NextGen atmosphere (Hauschildt et al. 1999) Cousins I and 2MASS *K* magnitudes from the tables given by France Allard on her web pages (http://phoenix.ens-lyon.fr/Grids/NextGen/). The NextGen Cousins *I* bolometric corrections that we use are from a more recent calculation than those used for the isochrones shown by J09. The difference is that the newer magnitudes are 0.34 brighter than the old. This shifts the Baraffe et al (2002) isochrone to the left of the majority of points in the color –magnitude diagram (CMD) shown by J09. When needed, conversions to the CIT system are made by applying transformations given by Carpenter (2001). In figure 2a, we compare the absolute Cousins *I* magnitudes for a solar composition star of solar luminosity and log $g$ = 5 as a function of $T_{eff}$. Fig 2b compares the absolute CIT *K* magnitudes.

The most luminous J09 data point corresponds to stars with $T_{eff} \approx 4000$ K. At this temperature, the Bessell et al. Cousins *I* magnitude is 0.21 larger than the NextGen Cousins *I* magnitude, with the VandenBerg & Clem magnitude lying between them. At this temperature, the NextGen and Bessell et al. *K* magnitudes are similar. Although there is evidence that the Victoria-Regina model isochrones, which use the VandenBerg & Clem colors, match observational color–absolute magnitude diagrams of nearby K–M dwarfs with precise Hipparcos parallaxes better than other isochrones (e.g. Bartašiūtė et al. 2012), we have opted not to use the VandenBerg & Clem results because of the lack of a *K* magnitude counterpart to their *I* magnitude. To get a measure of the uncertainties in our analysis that might result from differences in the adopted color and magnitude transformations, we use both the Bessell, Castelli & Plez (1998) and NextGen transformations.

From figure 2a, we see that near $T_{eff}$ = 4000 K a difference of 100 K between our Eddington boundary condition and atmospheres boundary condition models changes the *I* band magnitude by at most 0.05, which is much smaller than the 0.2 magnitude difference between the Bessell et al. (1998) and Hauschildt et al. (1999) bolometric corrections. At $T_{eff}$ = 3000 K, which is near the lowest temperature relevant to the J09 data, the effects of a 100 K change in $T_{eff}$ are larger but still smaller than the difference between the various determinations of the bolometric correction. From figure 2b, we see that, over the relevant temperature range, a change in $T_{eff}$ of 100 K changes the *K* magnitude by at most 0.05. The largest change in *I* – *K* color, of about 0.1, is less than the differences between the various color determinations. We conclude that the differences in bolometric corrections and colors between models using the Eddington approximation boundary conditions and models using the atmospheres boundary conditions do not have a significant effect on our results.

## 3. APPLICATION TO NGC 2516: MATCHING MODELS TO THE COLOR – MAGNITUDE DIAGRAM

To enable us to compare our theoretical results to the observations of J09 we have digitized the CMD shown in their figure 2. To superimpose the theoretical 158 Myr solar metallicity isochrone of Baraffe et al. (2002), J09 used a distance modulus for NGC 2516 of 7.93, which is slightly larger than the Hipparcos value of 7.68 ± 0.07 (Van Leeuwen 2009). Here we have decided to use the J09 distance modulus, because it leads to better agreement between our theoretical radii and the mean radii determined by J09. To account for reddening, we add the reddening correction of 0.213 used by J09 to the theoretical *I* magnitude and, based on the ratio of reddening in the *I* and *K* bands determined by Alves et al (2002), we also redden the theoretical *K* magnitude by 0.038.

### 3.1. Models with spots only: no inhibition of convection

We begin applying our models with spots but no magnetic inhibition of convection to the J09 CMD. J09 found that their measured mean radii are broadly consistent with a simple spot model in which 50% of the star's surface is covered by spots that are 30% cooler than the unspotted surface. To allow comparison with our results, we note that the equivalent coverage by completely dark spots corresponds to $f_s$ = 0.38. In figure 3, we show superimposed on the NGC 2516 CMD our 160 Myr isochrones for a range of values of $f_s$. In the left and right panels we use the BCP and NextGen bolometric corrections, respectively.

Focusing on the black lines which show the isochrones for stars without spots, we see that there are differences depending on which set of bolometric corrections is used. For the BCP set, the isochrone lies to the left of the bulk of data points for stars for which $R \sin i$ has been measured but at higher luminosities there a number of 'other members' which lie to the left of the isochrone. In contrast, the NextGen isochrone lies to the left of almost all the data points. We also see, for the more massive models, that adding spots can move the isochrones to the left of the isochrone for stars without spots. That this is possible is a result of the higher mass stars being not fully convective. Adding spots reduces the luminosity considerably less than for fully convective stars. For completely dark spots, the temperature of the radiating surface increases significantly and the isochrones move to the blue. Even with a spot contrast of 70%, the temperature of the unspotted surface increases significantly enough that the isochrones move to the blue. For the lower mass fully convective stars adding spots causes the luminosity to decrease, which for spot contrast of 70% decreases the temperatures of the both the unspotted surface and the spots. For spots that are not completely dark, the $I – K$ color generally lies between those calculated using the temperature of the unspotted surface, $T_u$, and the star's $T_{eff}$.

The main conclusion that we draw from figure 3 is that, even under the most favorable assumptions, at a given magnitude, the positions of the reddest stars in the J09 CMD would require spots coverage of more than 85% for spots cooler than the unspotted surface by 30%.

In the context of such huge spot coverages, we wonder: how likely is it that spot coverages greater than 85% are actually present in the stars of NGC 2516? To address this, we note that the photometric data of Irwin et al. (2007) indicate amplitudes of rotational modulation which range from roughly 0.4% to 5% for the more massive stars (0.4 – 0.7 $M_\odot$), with a median amplitude of about 1%. The range is from 0.7% to 7% for the less massive stars (0.2 – 0.4 $M_\odot$), with a median amplitude of about 1.5%. The central question now is the following: is it possible to convert, in a straightforward way, a photometric amplitude of rotational modulation to a fractional area of spot coverage? Unfortunately, the answer is in many cases "No". The amplitude of a star's photometric modulation actually traces the *asymmetry* in the star's spot distribution. In this regard, as far as the *longitude* of the spots is concerned, if the spots are distributed somewhat evenly in longitude, then the amplitude of rotational modulation could be small, even if the total areal coverage of the spots is large. Thus, axisymmetry in spot distribution is one way to produce small amplitude rotational modulation (Jackson & Jeffries 2012). Moreover, as far as the *latitude* of the spots is concerned, any spot which lies at a latitude in excess of the inclination $i$ of the rotation axis to the line of sight will always remain visible to the observer, and therefore result in no significant rotational modulation in the photometry. In order to convert photometry to spot distribution and areal coverage, a necessary piece of information is the value of $i$. This information is available typically only if the star in question is a member of a binary. Examples of low-mass binaries for which such information *is* available are the two low-mass spotted stars CC Eri and BY Dra of spectral type K7e (Bopp & Evans 1973): both have rotational periods of no more than a few days. With inclination information available (inclination $i \approx 30°$ and 40° for BY Dra and CC Eri, respectively [Bopp & Evans 1973]), these systems can be subjected to more reliable spot modeling. However, even in this case, the results are not complete: if there were to be a spot surrounding the pole, it would not contribute to modulation. As a result, the coverages derived from light curve modeling are lower limits. In CC Eri and BY Dra, the rotational modulations are observed to have amplitudes which vary from one observing season to another, with amplitudes which can be as small as 11% in some seasons, and as large as 30% in others. Modeling of the spots, using the constraints available from the binary orbits, indicates that the spots are (i) confined to low/moderate latitudes, and (ii) have fractional areas in the range 4 – 20%. As mentioned above, these are lower limits. In the case of CM Dra, consisting of 2 M dwarfs in a 1.3 day orbit, the areal coverage of spots is estimated to be as large as 17% (Morales et al. 2010). Thus, firm evidence already exists that certain cool dwarfs in binaries have spot coverages at least as large as 20%. On the other hand, CM Dra differs from BY Dra and CC Eri as regards latitudinal distribution of the spots: in CM Dra, the spots are mainly in the polar caps, and the longitudinal asymmetry of spot distribution is so small that the resulting photometric amplitudes are at most a few percent (Morales et al. 2010).

Unfortunately, the binary orbit approach is not immediately applicable to the stars in NGC 2516: J09 deliberately excluded the obvious binaries from their sample (see Section 3, first paragraph, in their paper). As a result, information about the inclination of rotational axes to the line of sight is not readily available for the J09 stars. Without such information, we cannot derive reliable spot coverages based solely on the photometric amplitudes of rotational modulation in NGC 2516: even though their median amplitudes are only 1 – 1.5% (i.e. 10 times smaller than the amplitudes in BY Dra or CC Eri), we cannot necessarily infer that spot coverages are smaller than those in CC Eri and BY Dra for the stars in NGC 2516.

In fact, the rotational properties of the stars suggests that spot coverages in NGC 2516 stars might not be small compared to those in BY Dra and CC Eri. For BY Dra and CC Eri, the values of $v \sin i$ are 5 and 15 km/sec (Bopp & Evans 1973). For the J09 stars with radius estimates in NGC 2516, $v \sin i$ values are > 8 km s$^{-1}$. As a result of this fast rotation, chromospheric activity in the NGC 2516 stars is high enough (Jackson & Jeffries 2010) to be at saturation levels for most of the stars in the sample: among the M dwarfs, all show saturated chromospheric emission. In view of this saturated behavior, there is no strong reason to expect very different levels of magnetism among the NGC 2516 stars. As a result, most or all of the J09 stars might be comparable in dynamo effectiveness to the dynamos which are operative in in BY Dra and/or CC Eri. Thus, NGC2516 stars *might* have spots with areal coverages as large as those on BY Dra and CC Eri (i.e. at least 4 – 20%), despite the small photometric amplitudes.

Independent evidence for large areal coverages of cool spots in a number of active dwarfs is provided by TiO data (O'Neal et al. 2004). In EK Dra (G1.5V), the minimum and maximum areal coverages at different rotational phases are 25% and 40%. In EQ Vir (K5Ve), the corresponding numbers are 33% and 45%. In a study of OH lines, O'Neal et al. (2001) found areal coverage of spots on LQ Hya (K2V) of 45±3%. Thus, in these very active G and K dwarfs, spot coverages rise to values of almost 50%.

Doppler imaging provides another line of evidence for spot coverages on rapidly rotating K and M dwarfs. For example, in a K dwarf with $v \sin i = 69$ km s$^{-1}$, there is evidence for low-latitude spots with diameters of up to 10 – 20°, and a polar spot which reaches down in some places to latitudes as low as 60° (Lister et al. 1999). In an M dwarf which also rotates at $v \sin i = 69$ km s$^{-1}$, there are spots at low and intermediate latitudes, but no evidence for polar spots (Barnes et al. 2004). The total fractional surface area of spot coverage in the Doppler images is estimated (by inspection) to be at most 10 – 20%, but these are lower limits due to limitations of Doppler imaging techniques.

Thus, despite the small photometric amplitudes of the M dwarfs in the J09 sample, we cannot exclude the possibility that those M dwarfs *might* have spot areal coverages of at least 10 – 20%. Moreover, if the areal coverages reported above for active G and K dwarfs can be extended to active M dwarfs, then we might not be able to exclude the possibility that spot coverages of 50% or more could be present in active M dwarfs.

In the present paper, our goal is to ask: can the J09 data be fitted with a model in which we can avoid the necessity of large spot coverages? We shall find that some of our models may allow us to do so (see Sections 3.2, 4.1, and 4.2 below).

*3.2. Models with inhibition of convection only; no spots*

Our conclusion from section 3.1 is that, rather than exploring models in which the empirical oversized radii of the low-mass stars in NGC 2516 are associated *solely* with the presence of spots, it would be worthwhile to explore a different type of model to explain the oversized radii of the stars in NGC 2516. In the model which we present here, we start by excluding spots altogether, and we concentrate in this Section on modeling the inhibition of convective onset in the limiting case where spots are absent. Subsequently (in Section 3.3) we will model the magnetic inhibition of convective onset while also allowing spots to be present, although not as large in fractional surface area as the coverages mentioned above.

Turning here to unspotted models, we report on results for models in which we incorporate a range of values of the magnetic inhibition parameter, $\delta$. In the presence of magnetic inhibition of convection, the radius of a star of given mass increases and its $T_{eff}$ decreases (MM01), and the larger the value of δ, the larger the radius and the lower the temperature become. The decrease in $T_{eff}$ makes a star of given mass redder than it would be in the absence of magnetic inhibition of convection.

In figure 4, we show the CMD plotted together with theoretical 160 Myr isochrones for non-magnetic stars (shown by black lines) together with the isochrones for δ values of 0.02, 0.04, 0.06, 0.08 and 0.09. In the left and right panels we use the BCP and NextGen bolometric corrections, respectively.

From the left panel, we see that when the BCP bolometric corrections are used, the data points for all but 6 of the stars with $R\sin i$ measurements lie in the region covered by the isochrones for values of the magnetic inhibition parameter, δ, between 0 and 0.08. However, a large number of the stars which are labeled "other members" lie outside this region. From the right panel we see all but 3 of the data points, including those for the "other members", lie in the region covered by the isochrones for values of δ between 0 and 0.09. The data point requiring the largest value of δ corresponds to a model of mass 0.34 $M_\odot$.

The conclusion of this section is that, in the absence of spots, many of the data of J09 *can* be fitted in the context of our (unspotted) magnetic inhibition model provided that magnetic fields corresponding to values of δ as large as 0.09 exist in the surface layers of the reddest stars in the J09 sample. In terms of surface magnetic field strength, a value of δ = 0.09 corresponds to a vertical field component of about 1270 G in a star where the surface pressure is 9.0 x $10^5$ dyne $cm^{-2}$. We return below (Section 3.4) to a discussion of the plausibility of the presence of fields of such strength on the surface of cool dwarfs.

However, one aspect of the unspotted model is inconsistent with J09 data: the model cannot account for rotational modulation, even though this is observed to be small (1-1.5%). This leads us to consider a hybrid model, to which we now turn.

*3.3. Models including both inhibition of convection and spots*

We now turn to cases which include not only magnetic inhibition of convection, but also allow for the presence of spots on the surface. For simplicity, we adopt a linear relationship between the spot coverage $f_s$ and the magnetic inhibition parameter δ. It seems reasonable that both $f_s$ and δ will be zero in the absence of any magnetic field. Hence we need only to determine the slope of the relationship. To do this we make use of our findings for CM Dra given in MacDonald & Mullan (2012). From fig. 15 of that paper, we see that the observed properties of CM Dra are consistent with a wide range of values of $f_s$ and δ. Again for simplicity, we use parameters near the center of these ranges, which lead to our adopted relationship

$$f_s = 17\delta. \tag{4}$$

The results of these hybrid spot – magnetic inhibition models are shown in figures 5a and 5b. A comparison with figures 3 and 4 shows similar trends. Even though the models are not all fully convective, their outer convection zones occupy a large volume of the stars which results in near degeneracy in $f_s$ and $\delta$. From the left panel of figure 5, we see that when the BCP bolometric corrections are used, the data points for all but 5 of the stars with $R\sin i$ measurements lie in the region covered by the isochrones for values of $\delta$ between 0 and 0.04. However, a large number of "other members" lie outside this region. From the right panel, we see that all but 1 of the data points, including those for the "other members", lie in the region covered by the isochrones for values of $\delta$ between 0 and 0.05. The reddest stars correspond to $\delta = 0.042 – 0.045$, $f_s = 0.71– 0.77$ for masses $0.26 – 0.36$ $M_\odot$. It appears that in our hybrid models, the data for the reddest stars in the J09 sample cannot be fitted unless we allow spot coverages to grow to values in excess of 70%. Once spots are included in the model, it is difficult to avoid the conclusion that the spots must have very large areal coverages in order to replicate the J09 results. And to replicate the observed rotational modulations of only $1.0 – 1.5\%$, the spots must be small and have statistically similar distributions in longitude. The vertical surface field component is ~ 900 G.

*3.4. Magnetic Field Strengths*

Eq. (0) allows us to determine the vertical component of the magnetic induction, $B_v$, at any point of our models. In particular, our models enable us to calculate the value of $B_v$ at the photosphere. For each data point in the portion of the CMD spanned by our hybrid model isochrones we have determined the corresponding mass and magnetic parameters (i.e. $\delta$ and/or $f_s$) by interpolation in our model results. In figure 6a, we show the photospheric $B_v$ values as a function of mass for the models *with* magnetic inhibition and spots, when the NextGen transformations have been used to find the magnitude and color. To compare with the results for models with magnetic inhibition but *without* spots, and also models in which the BCP corrections are used, we show in figure 6b the mean $B_v$ values obtained by binning in mass. The mass bin sizes are determined by having the same number of stars in each bin. From figure 6a, we see that the stars in the J09 sample are determined to have masses in the range $0.23 – 0.67$ $M_\odot$, and there is a general trend of decreasing magnetic field strength with increasing stellar mass. From Figure 6b, we see that models using the NextGen transformations predict higher magnetic fields than those using the BCP transformations, and higher magnetic field strengths are required for models without spots.

Figure 6a suggests that the magnetic fields of stars in NGC 2516 tend to zero for masses greater than about 0.7 $M_\odot$. But for less massive stars, our results suggest that magnetic fields contribute to the photometric properties of the stars of NGC 2516. There are several ways in which magnetic fields could affect the photometry: (i) the mere presence of spots on the surface will have an effect on the $T_{eff}$ of the "star" as viewed from afar; (ii) even if the star has no spots, magnetic fields may inhibit the onset of convection throughout the convection zone, thereby affecting the structure of the star; (iii) a combination of spots and magnetic inhibition. We find that for the magnetic inhibition models with or without spots, the mean surface vertical field strength is inversely correlated with mass for masses less than about 0.6 $M_\odot$. For the combined spot – magnetic inhibition model, the upper bound on the averaged surface vertical field values reaches values of about 600 G for stars of mass about 0.25 $M_\odot$, i.e. the masses that correspond to the lowest luminosity stars in the J09 sample. For the unspotted magnetic inhibition models the required surface vertical field strength could be as large as about 900 G for the lowest luminosity stars in the J09 sample.

Is there any observational evidence for vertical fields of such strengths on the surface of low-mass stars? We cannot answer this question specifically for M dwarfs with ages as young as NGC 2516. However, as regards M dwarfs in the solar neighborhood, Morin et al (2010) report that field stars with masses in the range 0.2 – 0.4 $M_\odot$ have poloidal fields with strengths in the range 400 – 800 G, while some stars in a low mass sample, with $M \leq 0.2$ $M_\odot$, have poloidal field strengths which can be as large as 1 – 2 kG. The ages of the field stars in the Morin et al sample are not known: but since the sample includes flare stars, and since flaring activity dies off with increasing age, it is at least possible that some of the stars in the Morin et al. sample may overlap in age with the stars of NGC 2516. To the extent that this is true, our requirement of having vertical fields up to 1 kG on the surface of a low-mass M dwarf is not inconsistent with observational evidence.

## 4. APPLICATION TO NGC 2516: MATCHING MODELS TO MEAN STELLAR RADII

The main result of J09 is that mean radii obtained by combining rotation periods with spectroscopic measurements of $v \sin i$ are up to greater than 50% larger than standard model radii. Here we compare radii from our models with spots and/or magnetic inhibition of convection with those obtained by J09. We first consider the radius – color relation.

### 4.1 *Radius – color comparison*

Once we have determined mass and magnetic parameters from the position of a star in the CMD, our models also provide the stellar radius. We have binned and averaged the predicted radii for those stars for which fits can be obtained with respect to the de-reddened *I - K* color. In figure 7, we show our average radii with 1$\sigma$ error bars together with the J09 results. From the CMD, we see that there is a large spread in luminosity at a given color. Assuming that the color is a reliable indicator of temperature, this spread in luminosity implies a large spread in radius. This is confirmed by the size of the error bars.

Figure 7a shows that 3 of the data points for the mean rotational radii are well matched by the theoretical mean radii for at least one combination of magnetic parameters. As can be seen from figure 7b, when the NextGen bolometric corrections are used all 5 rotational radii are matched by theoretical mean radii.

To quantify how well the models match the data, we give in table 1 values of

$$\eta_{col} = \frac{1}{5}\sum_{i=1}^{5}\frac{|R_i - R_{rot,i}|}{\sigma_{rot,i}}, \tag{5}$$

where $R_i$, $R_{rot,i}$ and $\sigma_{rot,i}$ are the mean theoretical radius, the rotational radius and its error, respectively, for the *i*th bin. We see that the best fit (i.e. smallest value of $\eta_{col}$) is obtained by the hybrid model with NextGen bolometric corrections. For this particular model, the mean spot coverages range from $f_s = 0.42$ for the bluest bin to $f_s = 0.63$. These spot coverage values are significantly less than for the spots only model for which $f_s$ ranges from 0.62 to 0.79. Since the pure magnetic inhibition model also gives a better fit than the spots only model, lower spot coverages may also be consistent with the radius data provided the magnetic inhibition parameter $\delta$ is appropriately increased.

*4.2 Radius – magnitude comparison*

When we compare radii as a function of absolute magnitude (see Figure 8), we find that the agreement between observation and theory is not as good as for the radius – color comparison. For the BCP bolometric corrections, only one mean rotational radius data point can be matched by theoretical mean radii. An improvement is found when the NextGen bolometric corrections are used. Now 2 rotational radii data points can be matched by theoretical mean radii, with near matches for two other data points. For the most luminous data point, the mean theoretical radii are found to be larger than the mean rotational radii. This may be in part due to the small number of objects in this bin. In particular we find for the spots only model that there are zero objects in this bin (which can also be seen from the color – magnitudes diagrams in figure 3).

By comparing to the radii of nonmagnetic models of the same luminosity, we see that almost all of the radius excess can be explained by including magnetic effects in the stellar models, particularly when the NextGen bolometric corrections are used. Again to quantify how well the models match the data, we give in table 1 values of $\eta_{mag}$ defined in the same way as $\eta_{col}$.

5. CONCLUSIONS AND DISCUSSION

Jackson et al (2009: J09) have determined mean radii of 147 low mass stars in the open cluster NGC 2516 for 5 bins each in $I - K$ color and absolute $I$ magnitude. In what follows, we refer to these J09 results as the "radius-color diagram" and the "radius-magnitude diagram". J09 find that the mean radii are significantly larger than predicted by standard stellar evolution models, and suggest that the radius discrepancies result from the presence of star spots. We have explored various kinds of magnetic stellar evolution models to see if we can fit the oversized radii. Among the models we explore, we include (i) spotted stars with no inhibition of convection, (ii) stars in which magnetic fields inhibit the onset of convection, but without any spots on the surface, and (iii) a hybrid model which combines the presence of magnetic inhibition with the presence of surface spots. We have applied these models to each star in the J09 color – magnitude diagram (CMD). We find that the pure spot model (i) is incapable of matching the positions of all of the stars in the CMD, particularly for stars with $I < 16$. However we find that the CMD positions of the bulk of the stars can be matched by pure magnetic inhibition models (ii) and by hybrid models (iii) that include the effects of both magnetic inhibition and star spots.

If we had used the older incorrect $I$ bolometric correction based on the NextGen atmosphere models, we would have found it difficult to match the CMD with any of our three magnetic models. The general effects of including magnetic inhibition and/or spots to models of fully convective or almost fully convective stars is to make them less luminous, redder and larger than models without magnetic effects. It is difficult to make low mass star models bluer by adding magnetic effects. A necessary requirement of our solution to the oversizing problem is that non-magnetic models should give isochrones that lie to the left of the CMD. This is the case when either the Bessell et al. (1998) bolometric corrections or the corrected NextGen-based bolometric corrections are used.

To compare with the mean radii of J09, we have binned our radius predictions in the same way as J09. In terms of the three types of models listed above, (i)-(iii), we find the following results.

(i) For models in which only the effects of spots are included, we find that the mean radii of the reddest stars in the sample would require that on average more than 79% of their surfaces be covered by completely dark spots. It is noteworthy that such huge spot coverages are also required, by observational constraints of rotational modulations, to cause only very small rotational modulations in the light levels of the NGC 2516 stars: the observed median modulations are only 1 – 1.5%. The spots have remarkably statistically similar distributions in longitude to produce such small rotational modulations despite covering more than 79% of the surface.

As an alternative to the presence of huge areal spot coverages, we have analyzed models of magnetic stars in which spots on the surface are not the *only* effect that the magnetic field causes in an M dwarf star: our models also include the inhibiting effects which a vertical magnetic field has on the onset of convection.

(ii) For models in which *only* the effects of magnetic inhibition of convection are included (without any spots on the surface), the trend in mean radii with mass among the J09 stars requires that the surface vertical field strength increases with decreasing mass for mass less than about 0.6 M$_\odot$. For the least massive stars, the needed vertical field strength $B_v$ in these unspotted models is approximately 650 – 900 G at the photosphere. We conclude that a magnetic inhibition model *can* replicate the J09 radius – color results without the need for surface spots as long as the surface fields are large enough. Surface field strengths of the requisite magnitudes are not inconsistent with observations of solar neighborhood stars (e.g. Morin et al. 2010). An inadequacy of unspotted models is their inability to explain the small amount (1.0 – 1.5%) of observed rotational modulation.

(iii) Finally, we have obtained "hybrid" models in which the combined effects of spots and magnetic inhibition of convection are considered. In these models, the required vertical field strength for the less massive stars in the J09 sample is lower: our models indicate that $B_v$ = 400 – 600 G at the photosphere. These specific hybrid models require that, in the reddest stars in the J09 sample, the areal coverage of the spots must be very large, with the average spot coverage in excess of 60%.

We are able to find models that fit the data in the radius – color diagram of J09 for models of types (ii) and (iii). Although none of our three models completely fit the data in the J09 radius- magnitude diagram, we find that magnetic effects can explain the bulk of the radius discrepancy. In view of the large spot areas associated with the specific type (iii) hybrid models, our preference is for either the type (ii) (unspotted) magnetic inhibition models or for a type (iii) model with lower spot coverage and larger magnetic inhibition.

Tables of data related to figures 3, 4 and 5 are provided in the online version of this paper.

ACKNOWLEDGMENTS
We thank R. D. Jeffries and an anonymous referee for helpful suggestions. JM thanks France Allard for help with the NextGen bolometric corrections and for updating the Cousins data, and Don VandenBerg for providing subroutines and data files for determining bolometric corrections. This work has been supported in part by the NASA Delaware Space Grant program and by a grant from the Mount Cuba Astronomical Foundation.

FIGURES

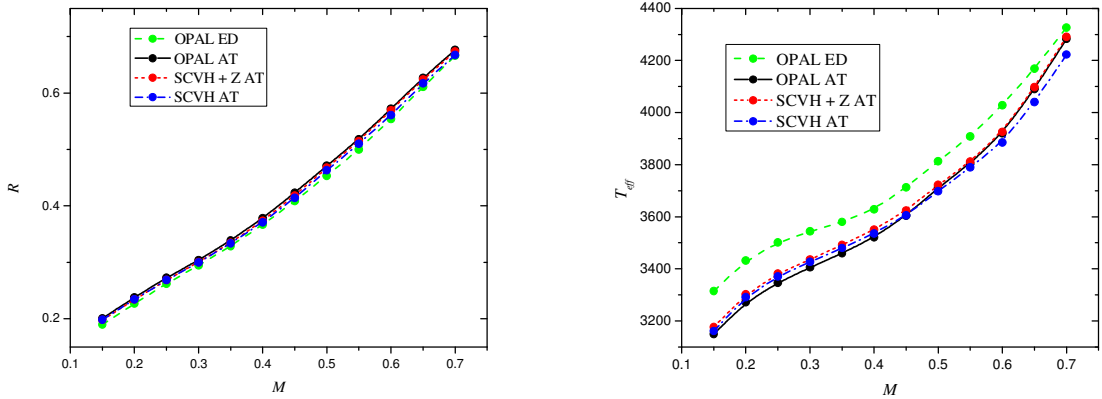

**Figure 1.** Comparison of solar composition 160 Myr radius – mass and $T_{eff}$ – mass isochrones for the OPAL and SCVH equations of states, and two different outer boundary conditions (see text for details). Note that the SCVH + Z eos is the SCVH eos modified to include the contribution from heavy elements. Mass and radius are in solar units, and $T_{eff}$ in Kelvin.

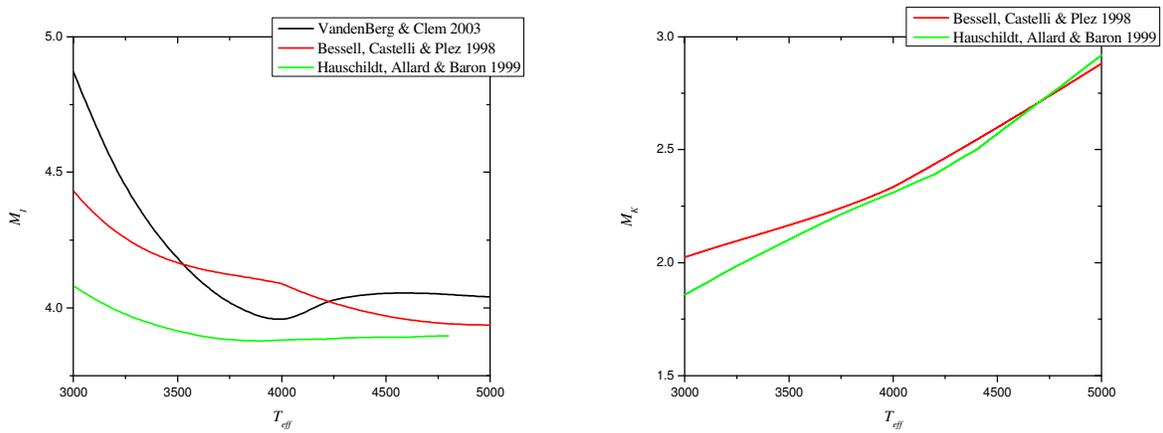

**Figure 2.** Comparison of the absolute $I_C$ and $K_{CIT}$ magnitudes for a 1 $L_\odot$, log $g$ = 5, solar composition star.

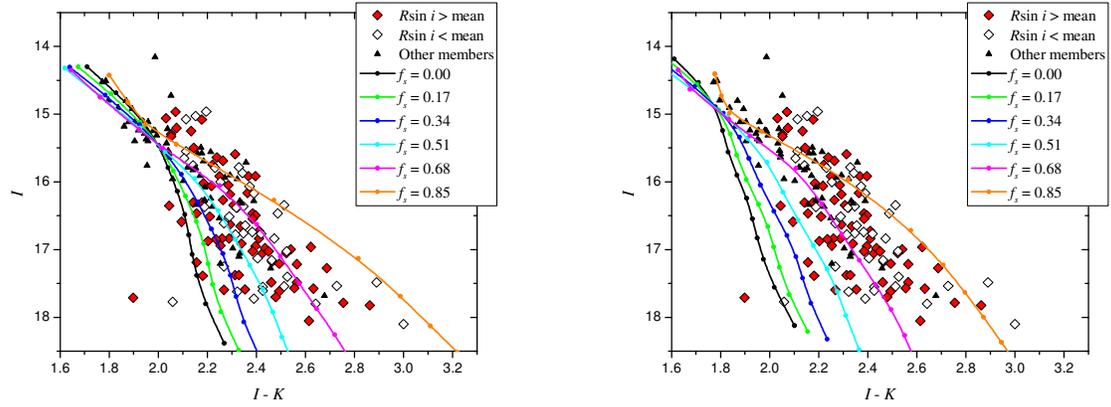

**Figure 3**. Comparison of 160 Myr isochrones for spotted stellar models with the CMD of NGC 2516 shown in figure 2 of J09. The left and right panels use the BCP and NextGen bolometric corrections respectively. The dark spots are assumed to have a temperature 70% of that of the unspotted surface. The filled circles on the isochrones indicate the mass points. The masses range from 0.7 $M_\odot$ to 0.15 $M_\odot$ in intervals of 0.05 $M_\odot$.

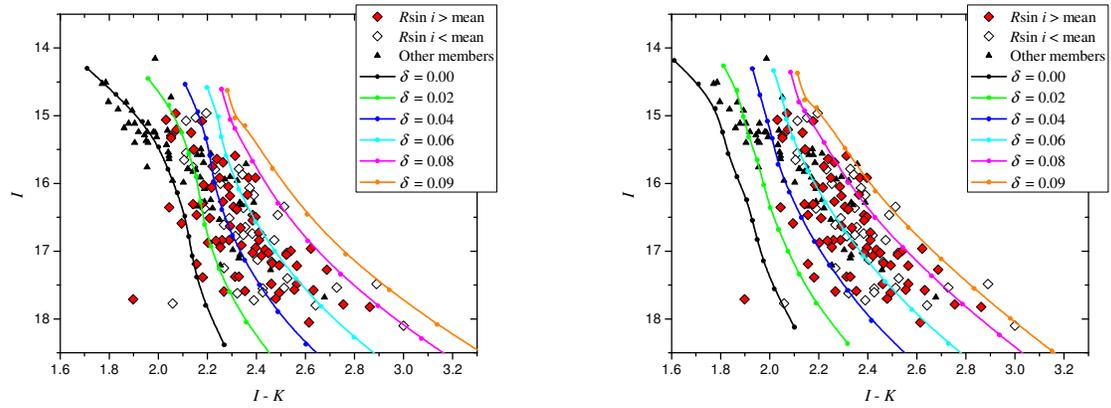

**Figure 4**. Comparison of 160 Myr isochrones for stellar models which include the inhibiting effects of magnetic fields on convection with the CMD of NGC 2516 shown in figure 2 of J09. The left and right panels use the BCP and NextGen bolometric corrections respectively. The filled circles on the isochrones indicate the mass points. The masses range from 0.7 $M_\odot$ to 0.15 $M_\odot$ in intervals of 0.05 $M_\odot$.

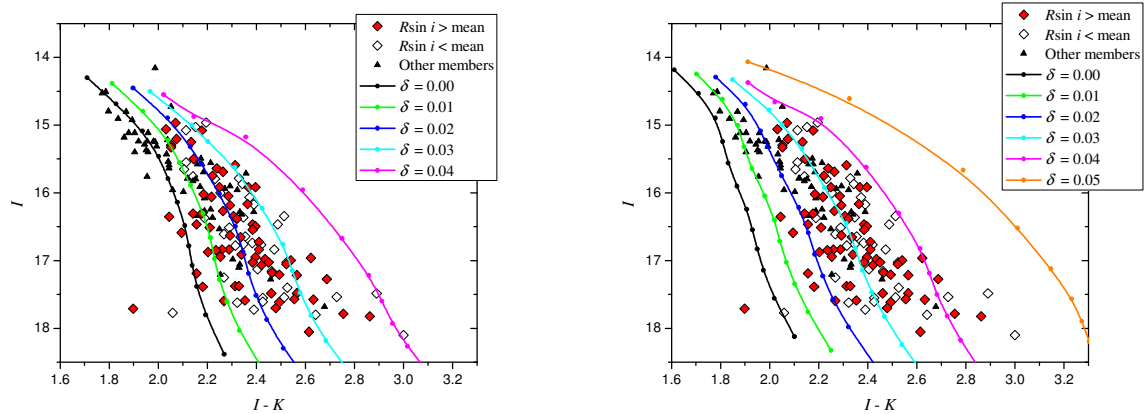

**Figure 5.** Comparison of 160 Myr isochrones for stellar models which include the effects of surface spots and magnetic inhibition of convection with the CMD of NGC 2516 shown in figure 2 of J09. The left and right panels use the BCP and NextGen bolometric corrections respectively. The dark spots are assumed to have a temperature 70% of that of the unspotted surface. The filled circles on the isochrones indicate the mass points. The masses range from 0.7 M$_\odot$ to 0.15 M$_\odot$ in intervals of 0.05 M$_\odot$.

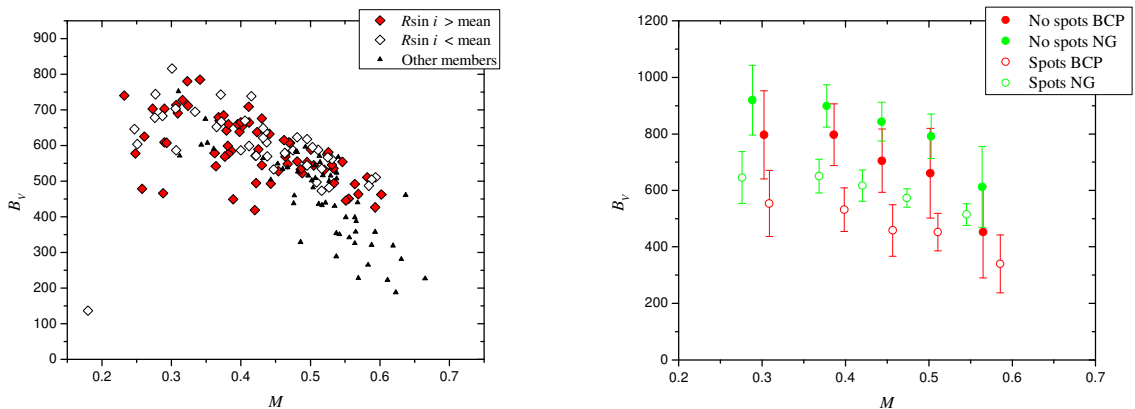

**Figure 6.** The left panel shows $B_v$, the vertical component of magnetic field at the photosphere, plotted against mass as determined from hybrid models applied to the individual data points in the J09 CMD assuming NextGen bolometric corrections. The right panel shows the average $B_v$ for when the data points are binned by mass. The mass $M$ is in solar units and $B_v$ is in units of Gauss.

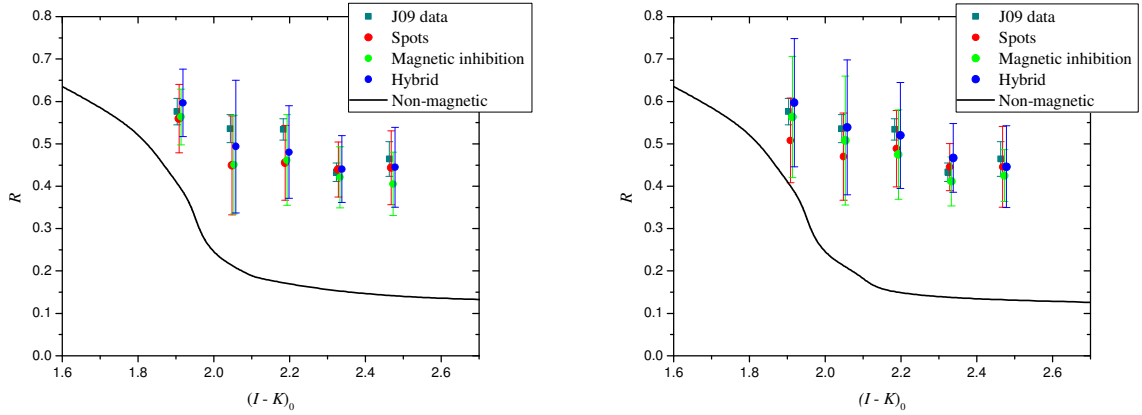

**Figure 7.** Average radii with 1$\sigma$ error bars for the three different sets of magnetic parameters compared with the rotational mean radii as function of de-reddened color. The left and right panels are for the BCP and NextGen bolometric corrections, respectively. For clarity the error bars have been offset by small amounts.

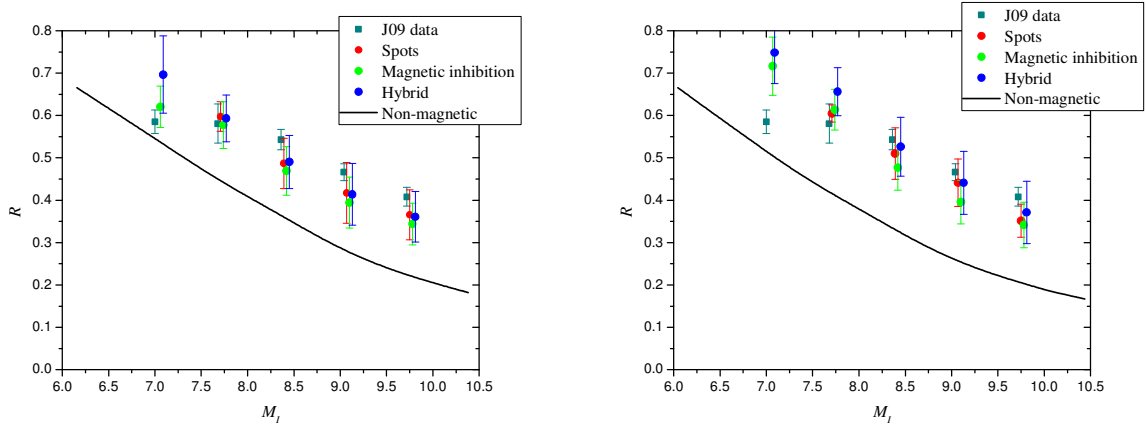

**Figure 8.** Average radii with 1$\sigma$ error bars for the three different sets of magnetic parameters compared with the rotational mean radii as function of de-reddened $I$ absolute magnitude. The left and right panels are for the BCP and NextGen bolometric corrections, respectively. For clarity the error bars have been offset by small amounts.

Table 1. Goodness of fit parameters for the radius-color and radius-magnitude diagrams. A smaller value indicates a better fit.

| Model | $\eta_{col}$ BCP | $\eta_{col}$ NextGen | $\eta_{mag}$ BCP | $\eta_{mag}$ NextGen |
|---|---|---|---|---|
| Spots | 1.42 | 1.41 | 1.77 | 1.46 |
| Magnetic inhibition | 1.57 | 1.12 | 2.19 | 2.94 |
| Hybrid | 0.98 | 0.66 | 2.24 | 2.22 |